\documentclass[amsmath,amssymb,amsfonts,superscriptaddress,reprint,showpacs,preprintnumbers,aps,prl,10pt]{revtex4-1}
\usepackage{xspace}
\usepackage[T1]{fontenc}
\usepackage{textcomp}
\usepackage{txfonts}
\usepackage{bm}

\usepackage{graphicx,color,epsfig}
\usepackage{fancyhdr}

\def\bbbc{{\mathchoice {\setbox0=\hbox{$\displaystyle\rm C$}\hbox{\hbox
to0pt{\kern0.4\wd0\vrule height0.9\ht0\hss}\box0}}
{\setbox0=\hbox{$\textstyle\rm C$}\hbox{\hbox
to0pt{\kern0.4\wd0\vrule height0.9\ht0\hss}\box0}}
{\setbox0=\hbox{$\scriptstyle\rm C$}\hbox{\hbox
to0pt{\kern0.4\wd0\vrule height0.9\ht0\hss}\box0}}
{\setbox0=\hbox{$\scriptscriptstyle\rm C$}\hbox{\hbox
to0pt{\kern0.4\wd0\vrule height0.9\ht0\hss}\box0}}}}

\newcommand{\ignore}[1]{}
\newcommand{\mComment}[1]{}
\newcommand{\gComment}[1]{}
\newcommand{\jComment}[1]{}
\newcommand{\rComment}[1]{}
\newcommand{\lComment}[1]{}
\renewcommand{\gComment}[1]{\textcolor{magenta}{Gerardo: #1}}
\begin{document}


\title{  Multiferroic behavior in trimerized Mott insulators }


\author{Y. Kamiya}

\author{C. D. Batista}

\affiliation{%
  Theoretical Division, Los Alamos National Laboratory, 
  Los Alamos, New Mexico 87545, USA
}%


\date{\today}

\begin{abstract}
  We demonstrate multiferroic behavior in trimerized Mott insulators through the interplay between spins and electric 
  dipole moments resulting from electronic charge fluctuations in frustrated units. 
  The model consists of stacked triangular layers of trimers with small intertrimer exchange interactions $J'$ and $J''$.
  Ferroelectric states coexist with ferro- or antiferromagnetic orderings depending on 
  the value of the magnetic field $H$ and the sign of the interlayer exchange $J''$. 
  The electric polarization undergoes abrupt changes as a function of $H$.
\end{abstract}

\pacs{%
  75.85.+t 
  71.30.+h 
  72.80.Sk 
  75.25.Dk 
}

\maketitle


{\it Introduction.}---%
Frustrated Mott insulators have been the focus of research  during the last decades~\cite{LMM2011Introduction}. 
The combination of geometric frustration with strong electron-electron interactions often leads to unusual collective behaviors. 
For instance, geometric frustration is a precondition for having nonuniform electronic charge distributions in the Mott phase 
of half-filled Hubbard models~\cite{Bulaevskii2008electronic}.  This requirement implies that the lattice must contain triangular units of magnetic ions. 
The simplest unit is an equilateral trimer of $S=1/2$ ions, like Cu$^{2+}$, whose ground states are two $S=1/2$ doublets.
This is true as long as the spin is a half-integer.  
Since the spin quantum number  is not enough to characterize each ground state, it is necessary to introduce an effective {\it orbital} degree of freedom (DOF) described by a  $\tau=1/2$ pseudospin variable~\cite{Bulaevskii2008electronic}. While $({\tau}_x, \tau_y)$ is proportional to the effective electric polarization operator, $ \tau_z$ is 
proportional to the  orbital magnetic moment produced by a current density that circulates around the trimer~\cite{Bulaevskii2008electronic}.

Materials that naturally include spin and orbital DOFs exhibit a variety of complex behaviors~\cite{Kugel1982Jahn-Teller}. 
In particular, orbital ordering often reduces magnetic frustration by creating disparities between effective exchange constants~\cite{Chern2011Spin}.  
While the $d$-orbital DOF of transition metals is related to electronic charge distributions with different quadrupolar moments,
the orbital DOF of  triangular molecules carries a net electric dipole moment.
Thus, collective behaviors of coupled trimers can lead to multiferroic phenomena or  magneto-electric effects arising from the interplay between the spin and orbital DOFs.

Trimers of magnetic ions are rather common in organic and inorganic compounds~\cite{Gudel1985Static,Honda1992Electron,Padilla1997Single-crystal,Sakurai2002Spin-half,Lopez2002Synthesis,Cage03,Nakano2005Synthesis,Stone2007Inelastic,Choi2008Pulsed-field}. 
They also exist in crystalline systems such as spin tubes, 
but the inter- and intratrimer exchange interactions are of comparable magnitude~\cite{Schnack2004Magnetic,Ivanov2010Heat,Manaka2011Effects}.
Although La$_4$Cu$_3$MoO$_{12}$\cite{Azuma2000Antiferromagnetic,Qiu2005Spin-trimer} is an ideal realization of weakly coupled trimers, the trimer superlattice does not favor a ferroelectric ordering~\cite{Wang2001Effective}. 
The advantage of organic environments is their flexibility for designing specific trimer lattices by choosing adequate ligand fields. 
While intramolecular exchange in frustrated molecules is a current topic of focus in magnetochemistry, little is known about the collective behaviors  induced by intermolecular couplings.

The purpose of this Letter is to bridge the gap between molecular and crystal magnets by demonstrating that multiferroic collective phenomena can arise from intertrimer exchange. 
After noticing that each trimer carries an internal electric dipole moment, 
it is natural to ask what is the effective coupling between these moments.
The answer depends on the nature of the trimer lattice and the sign of the exchange interaction~\cite{Wessel2001Phase,Wang2001Effective}. 
By demonstrating that a trimerized triangular lattice leads to ferro- or ferrielectric ordering, we provide guiding principles for designing new multiferroic materials.

We start by considering a Hubbard lattice of stacked triangular layers of trimers with small intertrimer hopping and large on-site Coulomb repulsion $U$. 
We find multiferroic ground states that remain stable up to magnetic fields above which the magnetization ($M$) and the electric polarization ($P$) change discontinuously.
These multiferroic states and strong magneto-electric effects are direct consequences  of the effective interaction between spin and orbital DOFs.

{\it Model.}---%
\begin{figure}[!b]
  \begin{center}
    \includegraphics[bb=0 0 346 623, angle=90, width=8.2cm]{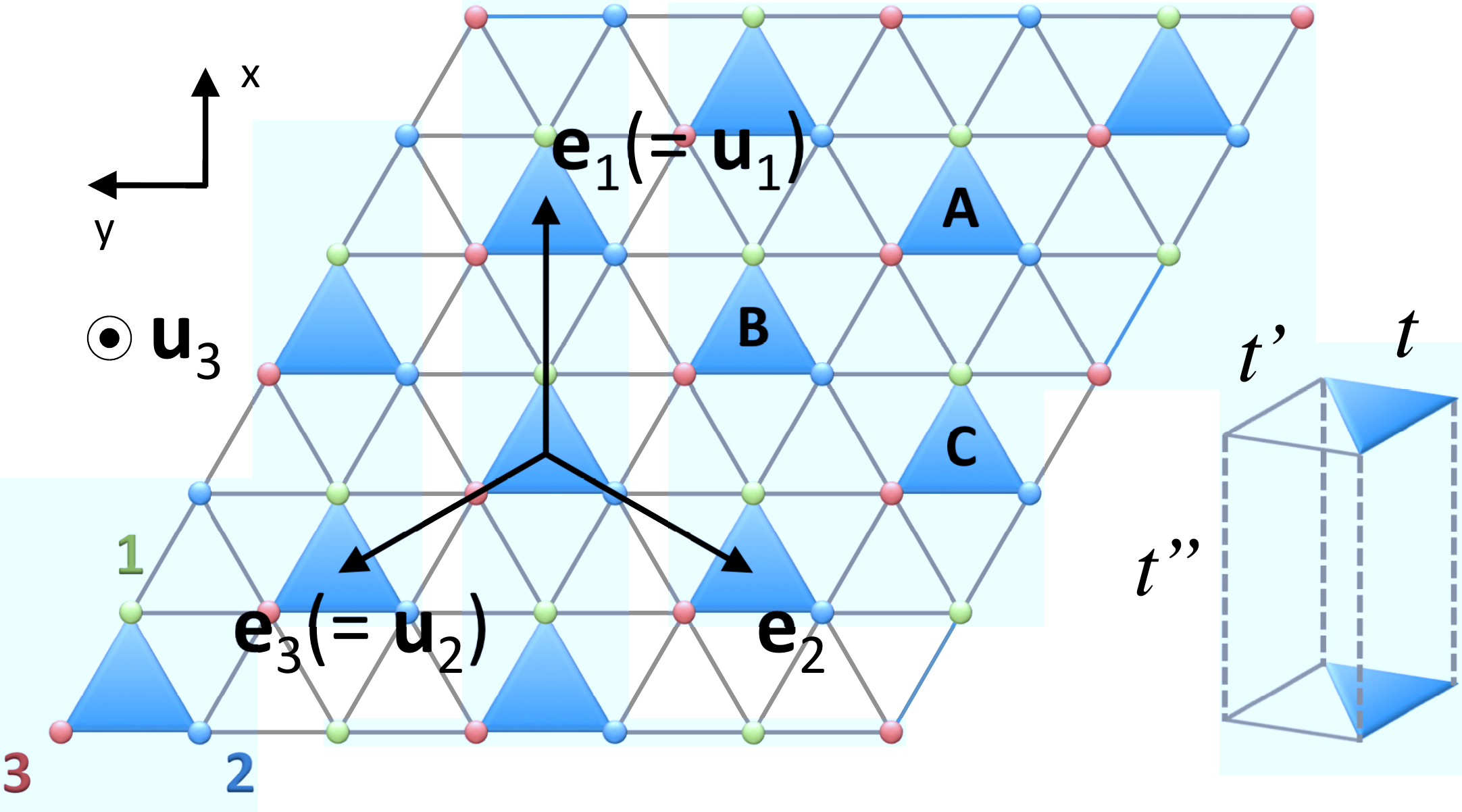}
  \end{center}
  \vspace{-0.5cm}
  \caption{%
    (color online). 
    Trimerized triangular lattice. 
    Shaded triangles represent trimers with dominant hopping amplitude. 
    The layers are stacked along the $c$ direction. $A,B,C$ are the indices for the three trimer sublattices.
  }
  \label{trimlat}
\end{figure}
The half-filled Hubbard model on the trimerized stacked triangular lattice of Fig.~\ref{trimlat} is
\begin{align}
  \mathcal{H} = -\sum_{{i}{j}\sigma}t_{{i}{j}}\left(c^{\dag}_{{i}\sigma}c^{\;}_{{j}\sigma} + {\rm{H.c.}}\right) + \frac{U}{2}\sum_{i}\left(n_i - 1\right)^{2},
\end{align}
where  $c^{\dag}_{{i}\sigma}$ ($c^{\;}_{{i}\sigma}$) is the creation (annihilation) operator of an electron with spin $\sigma$ at a site $i$, $n_{i} = \sum_{\sigma}c^{\dag}_{{i}\sigma}c^{\;}_{{i}\sigma}$ is the number operator, and the hopping amplitudes are $t_{ij} = t$ for $i$ and $j$ in the same trimer, and $ t_{ij} =t' (t'')$ when $i$ and $j$ are nearest-neighbor sites belonging to different trimers of the same layer (adjacent layers).
In what follows we consider the strong coupling limit $U \gg t$, though we will also comment on the intermediate-coupling regime $U \gtrsim t$.

When $U \gg t$, the half-filled Hubbard model can be reduced to a Heisenberg Hamiltonian ${\cal H}_\text{spin}$ by applying degenerate perturbation theory to the second order in $t_{{i}{j}}$:
\begin{multline}
  {\cal H}_\text{spin} = 
  J \sum_{{\bf r}, \mu > \eta} {\bf s}_{\mu, {\bf r}} \cdot {\bf s}_{\eta, {\bf r}}  
  + J' \sum_{{\bf r}, \eta, \mu \neq \eta} {\bf s}_{\eta, {\bf r}} \cdot {\bf s}_{\mu, {{\bf r}+{\bf e}_{\eta}}}  
  \\
  + J'' \sum_{{\bf r}, \eta} {\bf s}_{\eta, {\bf r}} \cdot {\bf s}_{\eta, {{\bf r}+ {\bf u}_3}} 
  - g \mu_B H \sum_{{\bf r}, \mu} s^z_{\mu, {\bf r}}.
  \label{eq:Hspin}
\end{multline}
Here we have added a Zeeman term to include the effect of an external field $H$ ($g$ is the gyromagnetic factor and $\mu_B$ is the Bohr magneton).
We have also refined our  notation by introducing the trimer coordinate, ${\bf r} = \sum_{i=1,2,3}n_i{\bf u}_i$, with ${\bf u}_i$ being primitive vectors for the trimer
lattice and  $\mu,\eta =1,2,3$ denoting the three ions of each trimer. 
${\bf e}_{\eta}$ are relative vectors between intralayer nearest-neighbor trimers (see Fig.~\ref{trimlat}).
${\cal H}_\text{spin}$ also describes systems of coupled spin tubes~\cite{Nishimoto2008Low-lying,Okunishi2005Low-Energy,Sakai2008Quantum,Sato2007Coexistence,Sato2007Vector,OkunishiArXivSpin}.

The reduction of ${\cal H}$ to a spin Hamiltonian for $U/t \gg 1$ suggests that only magnetic DOFs remain active at low energies.
However, virtual charge fluctuations can produce electric currents in loops or  electric dipoles ~\cite{Bulaevskii2008electronic}.
We will see below that this is indeed the case for the ground states of ${\cal H}$ as long as $t \gg t', t''$.
The exchange constants $J$, $J'$, and $J''$ are proportional to $t^2/U$, $t'^2/U$, and $t''^2/U$, respectively.
Therefore, spin trimers are weakly coupled ($J \gg J', J''$) for $t \gg t',t''$, and that will be the regime of interest from now on.
Although ${\cal H}$ leads to an antiferromagnetic (AFM) interlayer exchange $J''$, we will also consider the FM case, which can be realized for superexchange paths through  intermediate ions~\cite{Takahashi1991Discovery,Nakazawa1992Low-temperature,Shimizu2006Ferromagnetic}.

A single trimer has four degenerate ground states, namely, two $S=1/2$ doublets, that can be labeled as $\left\lvert{S^z,\tau^z}\right\rangle_{\bf r}$, 
where $S^z$ and $\tau^z$ are the ($\pm{1/2}$) eigenvalues of 
\begin{align}
  S^z_{\bf r} = s^z_{1,{\bf r}} + s^z_{2,{\bf r}} + s^z_{3,{\bf r}},~~
  \tau^z_{\bf r} = \frac{2}{\sqrt{3}}   {\bf s}_{1,{\bf r}} \times {\bf s}_{2,{\bf r}} \cdot {\bf s}_{3,{\bf r}}.
\end{align}
$\tau^z_{\bf r}$ is the scalar spin chirality that is proportional to the effective current density operator in the trimer ${\bf r}$. It  closes an SU(2) algebra with the operators,
\begin{align}
  \tau^x_{\bf r} &= \frac{1}{3} \left[2{\bf s}_{2,{\bf r}} \cdot {\bf s}_{3,{\bf r}} -{\bf s}_{1,{\bf r}} \cdot  ({\bf s}_{2,{\bf r}} + {\bf s}_{3,{\bf r}})\right]\;\;\propto{P_{\bf r}^x},
  \\
  \tau^y_{\bf r} &= \frac{1}{\sqrt{3}} {\bf s}_{1,{\bf r}} \cdot  ({\bf s}_{2,{\bf r}} - {\bf s}_{3,{\bf r}})\;\;\propto{P_{\bf r}^y},
\end{align}
which are proportional to the $x$ and $y$ components of the trimer electric dipole moment~\cite{Bulaevskii2008electronic}.
These spin and orbital trimer operators commute with each other: $[\tau^{\alpha}_{\bf r}, S^{\beta}_{\bf r'}] = 0$.

The $S=3/2$ excited states of the single trimer can be projected out as long as the single-trimer gap of $3J/2$ is much larger than $J'$ and $|J''|$.
The low-energy effective Hamiltonian is simply ${\cal H}_\text{eff} = {\cal P} {\cal H}_\text{spin} {\cal P}$, where ${\cal P}$ is the projector onto the subspace spanned by the direct product of single-trimer ground states $\left\lvert{S^z,\tau^z}\right\rangle_{\bf r} = \left\lvert{\pm{1/2},\pm{1/2}}\right\rangle_{\bf r}$~\cite{Subrahmanyam1995Block,Mila1998Low-Energy,Raghu2000Evaluation,Zhitomirsky2005Effective}.
${\cal H}_\text{eff}$ only includes contributions from interdimer interactions, which leads to
${\cal P} {\bf s}_{\mu, {\bf r}} \cdot {\bf s}_{\eta, {\bf r'}} {\cal P} =  {\cal P} {\bf s}_{\mu, {\bf r}} {\cal P} \cdot {\cal P} {\bf s}_{\eta, {\bf r'}} {\cal P}$. 
It is then convenient to introduce the projected spin operators:
\begin{equation}
  {\tilde {\bf s}}_{\mu,{\bf r}} \equiv {\cal P}{\bf s}_{\mu,{\bf r}}{\cal P} = \frac{1}{3}{\bf S}_{\bf r} \left[1 - 4{\boldsymbol \tau}_{\bf r}  \cdot {\bf n}_\mu\right],
  \label{eq:eff-spin}
\end{equation}
where
${\bf n}_{\mu} = \left(\cos\varphi_\mu, \sin\varphi_\mu, 0\right)$ with $\varphi_{\mu} = 4(\mu-1)\pi/3$ ($\mu = 1,2,3$)
are parallel to the displacement vectors from the center of the trimer to each spin site. 
${\tilde {\bf s}}_{\mu {\bf r}} $ must be proportional to $ {\bf S}_{\bf r}$ because this is the only vector under spin rotations in the single-trimer ground state subspace.
The proportionality factor cannot include  $\tau^z_{\bf r}$ because it is odd under  time reversal. This implies that ${\cal H}_\text{eff}$ will only include
 the $\tau^x$ and $\tau^y$ orbital variables associated with the local electric polarization of each trimer.
When the spin operators of Eq.~\eqref{eq:Hspin} are replaced by the projected operators of Eq.~\eqref{eq:eff-spin}, we obtain
\begin{eqnarray}
  {\cal H}_\text{eff} &=&  \frac{2 J'}{9} \sum_{{\bf r}, \eta}  
  {\bf S}_{\bf r} \cdot   {\bf S}_{{\bf r}+{\bf e}_{\eta}}  
  \nonumber \\
  &\times&
  \left[
    1 
    - 4 {\boldsymbol \tau}_{\bf r}  \cdot {\bf n}_{\eta} 
    + 2 {\boldsymbol \tau}_{{\bf r}+{\bf e}_{\eta} } \cdot {\bf n}_{\eta} 
    - 8  {\boldsymbol \tau}_{{\bf r}+{\bf e}_{\eta} } \cdot {\bf n}_{\eta} 
    {\boldsymbol \tau}_{\bf r}  \cdot {\bf n}_{\eta}
    \right] 
  \nonumber \\
  &+& \frac{J''}{3} \sum_{\bf r} {\bf S}_{ {\bf r} } \cdot {\bf S}_{ {\bf r} +{\bf u}_3}
  \left[
    1 + 8 ( \tau^x_{\bf r}  \tau^x_{{\bf r} + {\bf u}_3} + \tau^y_{\bf r}  \tau^y_{{\bf r} + {\bf u}_3} )
    \right]
  \nonumber \\
  &-& g \mu_B H \sum_{{\bf r}} S^z_{\bf r}.
  \label{eq:H-eff}
\end{eqnarray}
The continuous symmetry of the interlayer orbital coupling is a consequence of the $C_3$ lattice symmetry.

\begin{figure}[!t]
  \vspace{-0.375cm}
  \begin{center}
    \includegraphics[bb=0 0 584 655,angle=0,width=8.55cm]{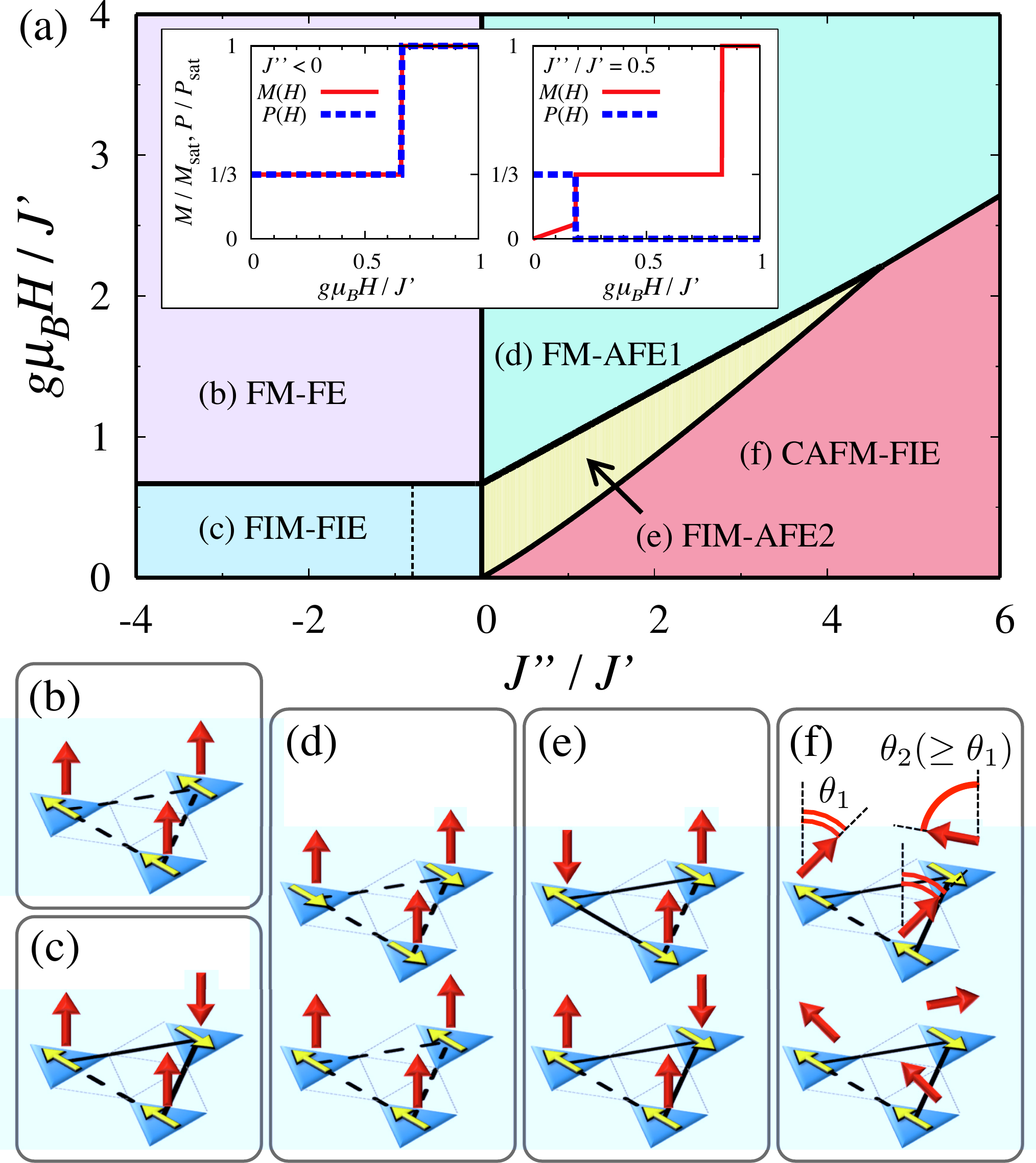} 
  \end{center}
  \vspace{-0.25cm}
  \caption{%
    (color online).
    (a) The MF phase diagram. The insets show $M(H)$ and $P(H)$ for $J'' < 0$ (left) and $J'' > 0$ (right).
    The dashed line shows an orbital transition due to the zero-point fluctuation.
    Also shown are schematic pictures of the MF solution [the arrows above (embedded on) trimers represent spins (orbitals)]: 
    (b) the FM-FE, (c) FIM-FIE, (d) FM-AFE1, (e) FIM-AFE2, and (f) CAFM-FIE states.
    The solid (dashed) lines indicate in-layer antiparallel (parallel) orbitals.
    Zero-field states are given up to an SU(2) global spin rotation.
  }
  \label{mean-field-states}
\end{figure}

{\it Above the saturation field.}---%
We will first consider the case of  magnetic fields that are large enough to polarize each trimer into the $S^z_{\bf r}=1/2$ state~\cite{PRL.trimer.comment1}: $H > H_{\text{sat}}$. 
Since ${\bf S}_{ {\bf r} } \cdot {\bf S}_{ {\bf r} + {\bf u}_3}={\bf S}_{ {\bf r} } \cdot {\bf S}_{ {\bf r} +{ {\bf e}_\eta}}=1/4$ holds in this case for the low-energy sector ${\cal S}_{\text{fp}}$ of ${\cal H}_\text{eff}$, the projection ${\cal P}_{{\cal S}_{\text{fp}}}$ of ${\cal H}_\text{eff}$ into ${\cal S}_{\text{fp}}$ leads to
\begin{eqnarray}
  {\cal H}'_\text{eff}  &\equiv& {\cal P}_{{\cal S}_{\text{fp}}} {\cal H}_\text{eff} {\cal P}_{{\cal S}_{\text{fp}}}= - \frac{ 4J'}{9} \sum_{{\bf r}, \eta}  
  {\boldsymbol \tau}_{{\bf r}+{\bf e}_{\eta} } \cdot {\bf n}_{\eta}  \;\;\;
  {\boldsymbol \tau}_{\bf r}  \cdot {\bf n}_{\eta}
  \nonumber \\
  &+& 
  \frac{2J''}{3} \sum_{\bf r} 
  \left(\tau^x_{\bf r}  \tau^x_{{\bf r} + {\bf u}_3} + \tau^y_{\bf r}  \tau^y_{{\bf r} + {\bf u}_3}\right) + \text{const}.
  \label{Hfm}
\end{eqnarray}
Here, the linear terms in ${\boldsymbol \tau}$ vanish because $\sum_{\mu}{\bf n}_\mu = 0$.
${\cal H}'_\text{eff}$ is an effective model for the orbital DOF which
 clearly shows that magnetic exchange interactions induce effective exchange couplings between the electric dipole moments. 
According to Eq.~\eqref{Hfm},  an AFM intralayer exchange, $J' > 0$, induces a ferroelectric (FE) exchange between electric dipoles in the same layer. 
In contrast, a FM (AFM) interlayer coupling results in a FE [antiferroelectric (AFE)] coupling along the $c$ direction.
We will now consider both possibilities by using a semiclassical approach and assuming that magnetic and orbital orderings are  three-sublattice structures. 
Since the effective interaction is {\it XY}-like, we will assume that $\langle \tau^z_{\bf r} \rangle=0$, $\forall {\bf r}$.
The {\it XY} components are determined by three variational parameters $\Phi_{l}$ with $l=A,B,C$ being the sublattice index of the trimers (see Fig.~\ref{trimlat}):
$\langle \tau^x_{\bf r} \rangle = (\sigma_{\bf r}/2) \cos{\Phi_l}$ and $\langle \tau^y_{\bf r} \rangle = (\sigma_{\bf r}/2) \sin{\Phi_l}$
for ${\bf r}\in{l}$. 
Here, $\sigma_{\bf r} = 1$ [$\sigma_{\bf r} = (-1)^{n_3}$ with $n_3$ being the layer index] for $J'' < 0$ ($J'' > 0$).
The mean-field (MF) energy per site that results from Eq.~\eqref{Hfm} is
\begin{equation}
  \epsilon_{H>H_{\text{sat}}} = -\frac{J'}{18}\sum_{l=A,B,C} \cos{(\Phi_l - \Phi_{l+1})} + \text{const}.
  \label{eq:eps-FP}
\end{equation}
It is clear from this expression that the global minimum of $\epsilon_{H>H_{\text{sat}}}$ is obtained for $\Phi_l = \Phi$.
This solution corresponds to a fully polarized FE (AFE) state for $J'' < 0$ ($J'' > 0$); see Figs.~\ref{mean-field-states}(b) and (d).
The arbitrary value of $\Phi$ implies that the MF energy is invariant under global orbital rotations along the $z$ axis; i.e., there is U(1) orbital symmetry at the MF level.

The next step is to introduce Holstein-Primakov bosons to describe the orbital fluctuations around the MF solution:
${\tau}^z_{\bf r} + i\bm{\tau}_{{\bf r}}\cdot\hat{\Omega}' = \sqrt{2\tau - n_{a{\bf r}}}\,a_{\bf r}$,
${\tau}^z_{\bf r} - i\bm{\tau}_{{\bf r}}\cdot\hat{\Omega}' = a^{\dag}_{\bf r}\sqrt{2\tau - n_{a{\bf r}}}$,
and
$\bm{\tau}_{{\bf r}}\cdot\hat{\Omega} = \tau - n_{a{\bf r}}$ for $J'' < 0$,
where $\hat{\Omega} = \left(\cos\Phi, \sin\Phi, 0\right)$, $\hat{\Omega}' = \left(\sin\Phi, -\cos\Phi, 0\right)$, $n_{a{\bf r}} = a^{\dag}_{\bf r} a^{\;}_{\bf r}$, and $\tau = 1/2$.
The case for $J'' > 0$ leads to the same ``orbital-wave'' Hamiltonian given below by a trivial gauge transformation. 
In terms of the Fourier modes, $a^{\;}_{\bf k}$ and $a^{\dag}_{\bf k}$, the Hamiltonian to order $1/\tau$ is
\begin{align}
  &{\cal H}^{\text{sw}}_{\tau, H>H_{\text{sat}}} = \sum_{\bf k} \left[ \epsilon_{\bf k} {a}^{\dagger}_{\bf k} {a}^{\;}_{\bf k}    
    + \frac{\gamma_{\bf k}}{2} \left({a}^{\dagger}_{\bf k}  {a}^{\dagger}_{-{\bf k}} +{a}^{\;}_{\bf k} {a}^{\;}_{-{\bf k}}\right)\right],
  \displaybreak[0]
  \notag\\
  &\epsilon_{\bf k} = \frac{2}{3}\left(J' - J''\right) - \gamma_{\bf k},
  \notag\\
  &\gamma_{\bf k} = \frac{2J'}{9}\sum_{\eta}\sin^2\left(\Phi - \varphi_\eta\right)\cos{\bf k}\cdot{\bf e}_{\eta} - \frac{J''}{3}\cos{k_z}.
  \label{eq:orbital-wave}
\end{align}
By performing a Bogoliubov transformation, we obtain
${\cal H}_{\tau,H>H_{\text{sat}}}^\text{sw} = \sum_{\bf k}[\omega_{\bf k} ( \alpha^{\dagger}_{\bf k}\alpha^{\;}_{\bf k}+ {1}/{2} ) - {\epsilon_{\bf k}}/{2}]$
with $\omega_{\bf k} = \sqrt{ \epsilon^2_{\bf k}  - \gamma^2_{\bf k} }$.
This dispersion relation has a zero mode at ${\bf k} = 0$ due to the U(1) invariance of the MF solution.
However, after including corrections due to zero-point fluctuations, the energy density,
${\epsilon}' = {\epsilon}_{H>H_{\text{sat}}} + \Delta\epsilon$ with $\Delta\epsilon = -(J' - J'')/3 + \sum_{\rm k}\omega_{\bf k}/2$, 
is minimized for $\Phi=(2n+1)\pi/6$, with $n$ being an integer number. 
Quantum fluctuations reduce the number of ground states according to the  sixfold symmetry of ${\cal{H}}'_\text{eff}$ and the maximum splitting between states with different values of $\Phi$ is of order $10^{-4}J'$. 
Consequently, higher order corrections in our $1/\tau$ expansion induce a very small gap in comparison to the exchange constants. 
For this reason, we will say that the orbital ${\bf k} = 0$ mode is a quasi-Goldstone mode.

{\it States for $H < H_\textit{sat}$.}---%
Now we consider the general case of an arbitrary magnetic field.
We propose a three-sublattice ordering in each layer and distinct even and odd layers for $J'' > 0$.
We use $(\theta_l^{b}, \phi_l^{b})$ to represent $\langle {\bf S}_{\bf r} \rangle$ and assume
$\langle \tau^x_{\bf r} \rangle = ({1}/{2})\cos{\Phi_l^{b}}$, 
$\langle \tau^y_{\bf r} \rangle = ({1}/{2})\sin{\Phi_l^{b}}$,
and
$\langle \tau^z_{\bf r} \rangle = 0$ for ${\bf r}\in{(b,l)}$ with $b = e, o$ and $l = A, B, C$. 
$\theta_l^{e} = \theta_l^{o}$, $\phi_l^{e} = \phi_l^{o}$, and $\Phi_l^{e} = \Phi_l^{o}$ are assumed for $J'' < 0$.
The MF energy density resulting from Eq.~\eqref{eq:H-eff} is
\begin{align}
  &\epsilon_\text{mf}
  = \frac{J'}{36}\sum_{b, l > l'}
  \left[
    \sin\theta_{l}^{b}\sin\theta_{l'}^{b}\cos\left(\phi_{l}^{b} - \phi_{l'}^{b}\right) + \cos\theta_{l}^{b}\cos\theta_{l'}^{b}
    \right]
  \notag\\
  &\times
  \left[
    1 - \cos\left(\Phi_{l}^{b} - \Phi_{l'}^{b}\right)
    \right]
  \notag \\
  &+\frac{J''}{36}\sum_{l}
  \left[
    \sin\theta^{e}_l\sin\theta^{o}_{l}\cos\left(\phi^{e}_l - \phi^{o}_{l}\right) + \cos\theta^{e}_l\cos\theta^{o}_{l}
    \right]
  \notag\\
  &\times
  \left[1 + 2\cos\left(\Phi^{e}_l - \Phi^{o}_{l}\right)\right]
  -\frac{g \mu_B H}{12}\sum_{b,l} \cos\theta_{l}^{b}.
  \label{eq:eps-H<Hc}
\end{align}
The phase diagram obtained by minimizing \eqref{eq:eps-H<Hc} is shown in Fig.~\ref{mean-field-states}(a).
As we already discussed, a FM-FE or FM-AFE1 state is stabilized for $H > H_{\text{sat}}$ depending on the sign of $J''$.
When $H < H_{\text{sat}}$, the MF ground state for $J'' < 0$ is the ferrimagnetic (FIM)-ferrielectric (FIE) state of Fig.~\ref{mean-field-states}(c) with an in-plane ``up-up-down'' orbital ordering that accompanies a corresponding up-up-down spin configuration,
e.g., $(\Phi_A, \Phi_B, \Phi_C) = (\Phi,\Phi,\Phi+\pi)$ and $({\bf S}_A, {\bf S}_B, {\bf S}_C) = (\uparrow,\uparrow,\downarrow)$ 
(a permutation of sublattice indices does not change the energy).
This state leads to 1/3 plateaus both in $M(H)$ and $P(H)$ 
that change discontinuously at $H = H_{\text{sat}}$ [see the inset of Fig.~\ref{mean-field-states}(a)]. 
The energy is minimized because the orbital ordering reduces the effective magnetic frustration.
This is the reason why the ground state differs from the $120^\circ$ structure for the spin-only AFM Heisenberg model on the triangular lattice.

On the other hand, the zero-field solution for $J'' > 0$ is the AFM-FIE state shown in Fig.~\ref{mean-field-states}(f) [the solution is given up to a global SU(2) spin rotation and $\theta_1 = \theta_2 = \pi/2$], which is also a collinear spin state.
In contrast to the $J'' < 0$ case, a finite $H$ induces the coplanar canted AFM (CAFM) state shown in Fig.~\ref{mean-field-states}(f), where $M(H)$ increases linearly while $P(H)$ remains constant at a 1/3 polarization plateau. 
Interestingly, the FIM-antiferroelectric (AFE2) state of Fig.~\ref{mean-field-states}(e) is stabilized for $J''/J' \lesssim 4.635$ in the intermediate field, leading to a 1/3 magnetization plateau as in the $J'' < 0$ case. 
$M(H)$ increases discontinuously while $P(H)$ vanishes abruptly at the first-order transition to this state.
The FIM-AFE2 state [see Fig.~\ref{mean-field-states}(e)] is such that in-layer FE bonds are oriented along different directions on adjacent layers.
This implies that there is an accidental extensive  degeneracy at the MF level that is removed by fluctuations. However,  the behavior of $M(H)$ and $P(H)$, namely, the large magneto-electric effect, is the same for all the degenerated states.
As for $J'' < 0$, the mechanism for stabilizing these states is a suppression of magnetic frustration by orbital ordering.
For any $J'' > 0$ the transition to the FM-AFE1 state [Fig.~\ref{mean-field-states}(d)] is of first order.

Finally, we include the spin-wave analysis for $H < H_{\text{sat}}$.
We will mainly discuss the $J'' < 0$ case (i.e., the FIM-FIE state). 
A more comprehensive analysis will be given elsewhere~\cite{KamiyaUnpublishedFullPaperTrimer}.
By choosing one of the FIM-FIE configurations, $(\Phi_A, \Phi_B, \Phi_C) = (0,0,\pi)$ and $({\bf S}_A, {\bf S}_B, {\bf S}_C) = (\uparrow,\uparrow,\downarrow)$, 
we introduce the Holstein-Primakov bosons, $b_{l{\bf r}}$ ($a_{l{\bf r}}$), for representing the spin (orbital) fluctuations on the sublattice $l$.
Spin and orbital fluctuations are decoupled in the linear spin-wave Hamiltonian.
Moreover, the orbital Hamiltonian can be reduced to Eq.~\eqref{eq:orbital-wave} simply by redefining $a_{\bf r} \equiv a_{l{\bf r}}$ for each ${\bf r}\in{l=A,B,C}$. 
This observation implies the stability of the FIM-FIE state against orbital fluctuations, and the presence of the quasi-Goldstone orbital mode.
The spin contribution of the spin-wave Hamiltonian is
\begin{align}
  {\cal H}_{S,\text{FIM-FIE}}^\text{sw}
  &= \frac{1}{2}\sum_{\bf k} 
  \left({\cal B}^{\dag}_{\bf k},{\cal B}^{\;}_{-{\bf k}}\right)
  \begin{pmatrix}
    P_{\bf k}^{\;} & Q_{\bf k}^{\;} \\
    Q_{-{\bf k}}^{\ast} & P_{-{\bf k}}^{\ast}
  \end{pmatrix}
  \left(
  \begin{array}{c}
    {\cal B}^{\;}_{\bf k} \\
    {\cal B}^{\dag}_{-{\bf k}}
  \end{array}
  \right)
  \notag\\
  &- N\left(\frac{g\mu_{B}H}{6} + \frac{4J'}{9} - \frac{J''}{2}\right),
\end{align}
where ${\cal B}_{\bf k} = \left(b_{A{\bf k}},b_{B{\bf k}},b_{C{\bf k}}\right)$ with $b_{l{\bf k}}$ being the Fourier transform of $b_{l{\bf r}}$.
$P_{\bf k}$ and $Q_{\bf k}$ are given by
\begin{align}
  P_{\bf k} = 
  \begin{pmatrix}
    d_{\bf k} & \gamma^{-}_{\bf k} & 0 \\
    (\gamma^{-}_{\bf k})^{\ast} & d_{\bf k} & 0 \\
    0 & 0 & d'_{\bf k}
  \end{pmatrix},~
  Q_{\bf k} = 
  \begin{pmatrix}
    0 & 0 & \gamma^{+}_{-{\bf k}} \\
    0 & 0 & \gamma^{+}_{\bf k} \\
    \gamma^{+}_{\bf k} & \gamma^{+}_{-{\bf k}} & 0
  \end{pmatrix},
\end{align}
where
$\gamma_{\bf k}^{\pm} = ({J'}/{9})
\sum_{\eta}\left[1 \pm 2\cos^2\left(\Phi - \varphi_\eta\right)\right] e^{i{\bf k}\cdot{\bf e}_{\eta}}$,
$d_{\bf k} = ({2J'}/{3}) - J''\left(1 - \cos{k_z}\right) + g\mu_B{H}$,
and
$d'_{\bf k} = d_{\bf k} + ({2J'}/{3}) - 2g\mu_B{H}$.
This Hamiltonian can be diagonalized by solving $\left\lvert\left(P_{\bf k} + Q_{\bf k}\right)\left(P_{\bf k} - Q_{\bf k}\right) - \omega_{\bf k}^2\right\rvert = 0$~\cite{Colpa1978Diagonalization} and a  positive-defined spectrum is obtained for any $H < H_{\text{sat}}$, indicating that the FIM-FIE state is locally stable. 
The spectrum is gapped for $0 < H < H_{\text{sat}}$.
This also implies stability of the 1/3 magnetization plateau for $J'' > 0$ (i.e., the FIM-AFE2 state) at least for sufficiently large $H$ and small $J''$ 
because both states are connected at $J'' = 0$.
Both spin and orbital fluctuations contribute to the zero-point energy that determines the precise structure of the orbital ordering.
We find that the effect of spin fluctuations competes against the effect of orbital fluctuations:
the energy minima of the spin contribution are $\Phi_\text{min}=n\pi/3$ in contrast to $\Phi_\text{min}=(2n+1)\pi/6$ for the orbital contribution.
The ground state energy turns out to be almost insensitive to $H/J'$ but it is significantly affected by $J''/J'$, which causes a phase transition from $\Phi_\text{min}=n\pi/3$ for the quasi-2D region $-0.8 \lesssim J''/J' (<0)$ to $\Phi_\text{min}=(2n+1)\pi/6$ for $J''/J' \lesssim -0.8$ [see Fig.~\ref{mean-field-states}(a)].

{\it Conclusions.}---In summary, the interplay between spin and orbital DOFs of weakly coupled trimers leads to multiferroic behavior and 
strong magneto-electric effects. Frustration plays a fundamental role in different stages of this problem. While it is crucial for the emergence of orbital DOFs that carry electric dipole moments~\cite{Bulaevskii2008electronic}, it is also  
essential for stabilizing the multiferroic orderings depicted in Fig.~\ref{mean-field-states}.  The  quasi-Goldstone orbital mode of these ordered states
can be indirectly observed  in the low temperature ($T$) regime by measuring its $T^{d}$ ($d$ is the spatial dimensionality) contribution to the specific heat  
for $T$ higher than the tiny orbital gap.  Spin excitations give a negligible contribution to the specific heat in the plateau phases because of the much larger spin gap.

Our perturbative  approach for weakly coupled trimers can be extended to the intermediate-coupling regime $U \gtrsim t$ as long as $|t| \gg t', t''$.
The effective spin-orbital Hamiltonian for $U \gtrsim t$ must be derived directly from ${\cal H}$ because ${\cal H}_\text{spin}$ no longer reproduces the low-energy spectrum of ${\cal H}$. 
The electric dipole of each trimer is much stronger in this regime (comparable to $e a$, where $e$ is the electronic charge and $a$ is the trimer lattice parameter) 
and can lead to polarizations of order 1$\mu$C/cm$^2$ 
if we assume a trimer density of 10$^{-3}$\AA$^{-3}$. 
Moreover, since the effective intertrimer coupling sets the scale of the ordering temperature $T_c$, it is possible to reach values of $T_c$ comparable or even higher than ambient temperature.

Quantum magnets comprising weakly-coupled dimers became the focus of intense research during the last two decades~\cite{Sebastian2006dimensional,Giamarchi2008bose}. 
We hope that the phase diagram presented in Fig.~\ref{mean-field-states} will trigger a similar effort in the search for realizations of weakly coupled trimers. 
We note that typical perturbations of triangular molecules, such as deviations from equilateral shape or intratrimer Dzyaloshinskii-Moriya interactions, will not induce important changes in
our phase diagram 
as long as they are small in comparison to $J'$ and $J''$. 
\begin{acknowledgments}
  We thank Y. Takano, H. Manaka, K. Okunishi, P. Jain, and S.-W. Choeng for valuable comments.
  Work at LANL was performed under the auspices of the
  U.S.\ DOE Contract No.~DE-AC52-06NA25396 through the LDRD program.
\end{acknowledgments}

\bibliography{references}

\end{document}